\documentclass[11pt]{article}
\usepackage{amsfonts,amssymb,amsmath}

\usepackage[english]{babel}
\newtheorem{theorem}{Theorem}
\newtheorem{definition}{Definition}

\newtheorem{proof}{Proof}

\newtheorem{proposition}{Proposition}
\newtheorem{lemma}{Lemma}

\newcommand{\beq}{\begin{eqnarray}}
\newcommand{\eeq}{\end{eqnarray}}

\newcommand{\beqt}{\begin{eqnarray*}}
\newcommand{\eeqt}{\end{eqnarray*}}

\newcommand{\be}{\begin{equation}}
\newcommand{\ee}{\end{equation}}

\newcommand{\bl}{\begin{lemma}}
\newcommand{\el}{\end{lemma}}

\newcommand{\bt}{\begin{theorem}}
\newcommand{\et}{\end{theorem}}

\newcommand{\bd}{\begin{definition}}
\newcommand{\ed}{\end{definition}}

\newcommand{\bp}{\begin{proposition}}
\newcommand{\ep}{\end{proposition}}

\newcommand{\bpr}{\begin{proof}}
\newcommand{\epr}{\end{proof}}

\newcommand{\bi}{\begin{itemize}}
\newcommand{\ei}{\end{itemize}}

\newcommand{\ben}{\begin{enumerate}}
\newcommand{\een}{\end{enumerate}}

\newcommand{\Z}{\mathbb Z}

\newcommand{\E}{\mathbb E}

\newcommand{\s}{\ensuremath{\mathcal{S}}}

\newcommand{\om}{\ensuremath{\omega}}
\newcommand{\Om}{\ensuremath{\Omega}}

\newcommand{\La}{\ensuremath{\Lambda}}
\newcommand{\si}{\ensuremath{\sigma}}

\begin{document}

\title{{\bf Decimation of the Dyson-Ising Ferromagnet}}

\author{Aernout van Enter\footnote{Johann Bernoulli Institute, University of Groningen, Nijenborgh 9, 9747AG,Groningen, Netherlands} \, and Arnaud Le Ny\footnote{LAMA UMR CNRS 8050, UPEC, Universit\'e Paris-Est,  94010 Cr\'eteil, France.}\footnote{E-mail: aenter@phys.rug.nl,  arnaud.le-ny@u-pec.fr}}

\maketitle

\begin{center}
{\bf Abstract:} 
%{\bf Preliminary Draft}}
\end{center}
We study the decimation to a sublattice of half the sites of the one-dimensional
Dyson-Ising ferromagnet with slowly decaying long-range pair potentials of the 
form $\frac{1}{| i-j |^\alpha}$,  deep in the phase transition region 
($1 < \alpha \leq 2$ and low temperature). We prove non-Gibbsianness of the 
decimated measures at low enough temperatures by exhibiting a point of essential
discontinuity for the (finite-volume) conditional probabilities of 
decimated Gibbs measures.
This result complements previous work proving conservation of Gibbsianness 
for fastly decaying potentials ($\alpha >2$) and provides an example of a 
"standard" non-Gibbsian result in one dimension, in the vein of similar results 
in higher dimensions for short-range models. We also discuss how these measures could fit
within a generalized (almost vs. weak) Gibbsian framework.  Moreover  we comment on the possibility of similar results for some other transformations.

\medskip

%\footnotesize

\vspace{7cm}



\newpage
\section{Introduction}

In this paper, we focus on  properties of transformed equilibrium measures of  one-dimensi\-onal Ising models with long-range, polynomially 
decaying, pair interactions called {\em Dyson-Ising models} or 
%sometimes shortly 
just 
{\em Dyson Models}. These models display a phase transition at low temperature, for appropriate values of the decay parameter. 
Varying this decay parameter plays a similar role as varying the dimension in 
short-range models. This can be done in a continuous manner, so one has  analogues of well-defined models in continuously varying non-integer dimensions, which is a major reason why these models have attracted a lot of attention in the study of phase transitions and critical behaviour (see e.g. \cite{CFMP} and references therein).
Here we show that,  at low enough temperature, under a decimation transformation the low-temperature measures of the Dyson models are mapped to non-Gibbsian measures, similarly to what happens for short-range interactions in higher dimensions. We also discuss possible extensions within the generalized Gibbs framework and some related issues.  
 
%-Failure of quasilocality or essential discontinuity, even exceptional (set of measure 0),  responsible for non-Gibbsianness, provocates a breakdown of the standard proofs behind the interpretation of Gibbs measures as equilibrium states and behind the understanding of the phase diagrams and the properties of its phases.\\

%General long-range, $J(n)$, $\sum_n J(n) < \infty$, etc. $J(n) \equiv n^{(-\alpha)}$\\

 %When the range of the interaction is lower, i.e. for $\alpha >2$, uniqueness holds and Redig and Wang \cite{RW} have proved that Gibbsianness was conserved, precising in some cases ($\alpha >3$) a decay of correlation for some transformed potential.

The paper is organized as follows. In Section 2, we describe the standard DLR approach to Gibbs measures in mathematical statistical mechanics --  including "global specifications"  \cite{FP} -- and our long-range Dyson-Ising models. In Section 3, we introduce the decimation transformation -- an elementary renormalization transformation that keeps odd or even spins only -- and prove non-Gibbsianness at low temperature for the decimated Dyson-Ising models whose interactions are so slowly decaying that, conditioned on the even spins to be alternating,  a  ``hidden phase transition'' occurs in the system of odd spins. Eventually, in Section 4, we extend previous results to show that  this decimated measure is included in the class of Almost Gibbsian measures, and comment on some related issues. 
\section{ Gibbs Measures, Background and Notation}
\subsection{Specifications and Measures}

We 
%thus 
will
deal  with long-range ferromagnetic Ising models 
with pair interactions in one dimension. These are part of the more general class of lattice (spin) models with Gibbs measures, as discussed for example in \cite{Fer,FV,HOG,VEFS}.
The finite-spin state space is the usual Ising space 
$(E,\mathcal{E},\rho_0)$ with 
$E=\{-1,+1\}$, $\mathcal{E}=\mathcal{P}(\{-1,+1\})$ and the a priori counting 
measure $\rho_0=\frac{1}{2} \delta_{-1} + \frac{1}{2} \delta_{+1}$. We denote by 
$\mathcal{S}$ the set of the finite subsets of $\Z$ and, for any $\La \in \s$, 
write $(\Om_\La,\mathcal{F}_\La,\rho_\La)$ for the finite-volume configuration 
space $(E^\La,\mathcal{E}^{\otimes \La},\rho_o^{\otimes \La})$. At infinite volume, 
%microscopic states are represented by 
configurations are denoted by $\si,\om$, etc., lying in an infinite-volume
 {\em configuration space}, the infinite-product probability space 
$(\Omega,\mathcal{F},\rho)=(E^\Z,\mathcal{E}^{\otimes \Z},\rho_0^{\otimes \Z})$, 
equipped with the product topology of the discrete topology on $E$.  For this 
topology, continuous functions coincide with {\em quasilocal} functions, that is, uniform limits of local functions, 
the latter being $\mathcal{F}_\La$-measurable functions for some $\La \in \s$. A
 function is said to be {\em right-continuous} (resp. {\em left-continuous}) 
when for every $\om \in \Om$, $\lim_{\La \uparrow \s} f(\om_\La +_{\La^c})=f(\om)$ 
(resp. $\lim_{\La \uparrow \s} f(\om_\La -_{\La^c})=f(\om))$, where one writes 
$\om_\La$ for its projection on $\Om_\La$, and $+$ (resp. $-$) for the configurations
 whose value are respectively  $+1$ (resp. $- 1$) everywhere. We also generically 
consider infinite subsets $S \subset \Z$, for which all the preceding notations 
defined for finite $\Lambda$ extend naturally 
($\Omega_S,\mathcal{F}_S, \rho_S, \sigma_S$, etc.). Important events to be 
considered are the {\em asymptotic events}, which are the elements of the 
{\em tail} $\sigma${\em-algebra} $\mathcal{F}_{\infty}=\cap_{\Lambda \in \mathcal{S}} \mathcal{F}_{\Lambda^c}$. These events typically do not depend on any local  behaviour, that is, they are insensitive to changes of any finite number of spins, and are mostly obtained by some limiting procedure. 

Within the  product topology, configurations are close when they coincide on large finite regions $\Lambda$, and the larger the region, the closer they are. For a given configuration $\omega \in \Om$, a  neighborhood base is thus provided by the family $\big(\mathcal{N}_\Lambda(\omega)\big)_{\Lambda \in \mathcal{S}}$ with, for any $\Lambda \in \s$,
$$
\mathcal{N}_\Lambda(\omega)=\Big \{ \sigma \in \Omega : \sigma_\Lambda=\omega_{\Lambda},\; \sigma_{\Lambda^c} \; {\rm arbitrary} \Big\}.
$$
We also consider particular open subsets of neighborhoods $\mathcal{N}_\Lambda(\omega)$ on which the configuration is $+$ (resp. $-$) on an annulus $\Delta \setminus \Lambda$ for $\Delta \supset \Lambda$, defined for all $\Lambda \in \s,\; \om \in \Om$ as
$$
\mathcal{N}_{\Lambda,\Delta}^+(\omega) = \Big \{ \sigma \in \mathcal{N}_\Lambda(\omega) : \sigma_{\Delta \setminus \Lambda } = +_{\Delta \setminus \Lambda},\; \sigma_{\Delta^c} \; {\rm arbitrary} \Big\} \; \big({\rm resp.} \; \mathcal{N}_{\Lambda,\Delta}^-(\omega)\big).
$$
% = \Big \{ \sigma \in\mathcal{N}_\Lambda(\omega) : \sigma_{\Delta \setminus \Lambda } = -_{\Delta \setminus \Lambda},\; \sigma_{\Delta^c} \; {\rm arbitrary} \Big\}.
%\end{eqnarray*}

We denote by $C(\Om)$ the set of continuous  functions on $\Om$. In our  finite state-space set-up, continuity is equivalent to uniform continuity and to  {\em quasilocality}\footnote{Continuous functions are uniform  limits of local functions, explaining the terminology {\em quasilocal} \cite{Fer,HOG}.}, so that one has
\be \label{qlocfu}
f \in C(\Omega) \; \Longleftrightarrow \; \lim_{\Lambda \uparrow \s} \sup_{\sigma,\omega:\sigma_\Lambda=\omega_\Lambda} \mid f(\omega) - f(\sigma) \mid = 0.
\ee

We also  will make at various points use of the existence of  a partial order 
(FKG)  $\leq$ on $\Omega$: $\sigma \leq \omega$ if and only if 
$\sigma_i \leq \omega_i$ for all $i \in \Z$. Its maximal and minimal elements 
are the configurations $+$ and $-$, and this order extends to functions: 
$f:\Omega \longrightarrow \mathbb{R}$ is called {\em monotone increasing}  when 
$\sigma \leq \omega$ implies $f(\sigma) \leq f(\omega)$. It induces then a 
stochastic domination on probability measures on $\Omega$ for which  we write $\mu \leq \nu$ if and only
if $\mu[f] \leq \nu[f]$ for all $f$ monotone increasing, where we denote 
$\mu[f]$ for the expectation $\E_\mu[f]$.
\\

%macroscopic 
 States are  represented by the  set  $\mathcal{M}_1^+$ of probability measures 
on the configuration space $(\Omega,\mathcal{F},\rho$). To describe such 
measures on the infinite product space $\Omega$ in a way that would {\em not} 
necessarily lead to uniqueness, and thereby allow to mathematically describe phase 
transitions, Dobrushin \cite{Dob1} and Lanford/Ruelle \cite{LaR} introduced in 
the late 60's an approach where a measure is 
%described by the prescription of 
required to have prescribed  
conditional probabilities w.r.t. the outside of {\em finite} sets. 
Such a system of conditional probabilities extended to be defined {\em everywhere}, rather than almost everywhere because one does not have yet a measure to begin with, is called a {\em specification}.
%\\
%**********************************************************************\\
%I reduced a bit the defintion to answer ref 2, ad also incorporate part of the global specification paragraph there, reducing a but its introduction. The other part -- the construction for Dyson -- has been left in the subsection of Dyson.\\
%***********************************************************************\\
%to represent such systems are the following  particular families of probability 
%kernels:
%\footnote{Formally introduced by F\"ollmer \cite{Foll0} and Preston \cite{Pr0} in the mid 70's.}:
\begin{definition}[Specification]:\\
A {\em specification} $\gamma=\big(\gamma_\Lambda\big)_{\Lambda \in \s}$  on $(\Omega,\mathcal{F})$ is a family of probability kernels  $\gamma_\Lambda : \Omega_\Lambda \times \mathcal{F}_{\Lambda^c} \; \longrightarrow \; [0,1];\; (\omega,A) \; \longmapsto \;\gamma_\Lambda(A \mid \omega)$
s.t. for all $\Lambda \in \mathcal{S}$:
\begin{enumerate}
%\item For all $\omega \in \Omega$, $\gamma_\Lambda(\cdot | \omega)$ is a probability measure on $(\Omega,\mathcal{F})$.
%\item For all $A \in \mathcal{F}$, $\gamma_\Lambda(A | \cdot)$ is $\mathcal{F}_{\Lambda^c}$-measurable.
\item (Properness) For all $\omega \in \Omega$, $\gamma_\Lambda(B|\omega)=\mathbf{1}_B(\omega)$ when $B \in \mathcal{F}_{\Lambda^c}$.
\item (Finite-Volume consistency) For all $\Lambda \subset \Lambda' \in \s$, $\gamma_{\Lambda'} \gamma_{\Lambda}=\gamma_{\Lambda'}$ where 
\be \label{DLR0}
\forall A \in \mathcal{F},\; \forall \omega \in \Omega,\;(\gamma_{\Lambda'} \gamma_\Lambda)(A | \omega)=\int_\Omega \gamma_\Lambda(A | \sigma) \gamma_{\Lambda'}(d \sigma | \omega).
\ee
\end{enumerate}
\end{definition}
These kernels also act on functions and on measures: for all $f \in C(\Omega)$ or $\mu \in \mathcal{M}_1^+$,
$$
\gamma_\Lambda f(\omega):=\int_\Omega f(\sigma) \gamma_\Lambda (d \sigma | \omega)=\gamma_\Lambda [f | \omega] \; {\rm and} \; 
\mu \gamma_\Lambda [f] : = \int_\Omega (\gamma_\Lambda f)(\omega) d \mu (\omega)= \int_\Omega \gamma_\Lambda [f | \omega] \mu(d \omega).
$$

%These objects are designed to represent consistent systems of conditional probabilities, with the important 
%objection 
%additional property
%that they are defined everywhere and not only almost-surely as ordinarily conditional probabilities would have been required to be. However, as we do not have a measure to begin with, the notion of "almost surely" a priori does not make sense.   

For a given specification, different measures can then have their conditional probabilities represented by the  same specification (and satisfy the {\em DLR equations} (\ref{DLR1})) but live on different full-measure sets. This leaves the door open to a mathematical description of phase transitions, which is well known e.g. for the ferromagnetic (n.n.) Ising model on the square lattice $\Z^2$ \cite{Grif}, but also for our long-range Ising models on $\Z$, see next section.

\begin{definition}[DLR measures]:\\
A probability measure $\mu$ on $(\Omega,\mathcal{F})$ is said to be consistent with a specification $\gamma$ (or specified by $\gamma$) when  for all $A \in \mathcal{F}$ and $\Lambda \in \s$
\be \label{DLR1}
\mu[A|\mathcal{F}_{\Lambda^c}](\omega)=\gamma_\Lambda(A|\omega), \; \mu{\rm -a.e.} \;  \omega.
\ee
%Equivalently, $\mu$ is consistent with $\gamma$ when $\mu=\mu \gamma_\Lambda$ for all $\Lambda \in \s$, i.e. when
%\be \label{DLR2}
%\int (\gamma_\Lambda f) d \mu = \int f d \mu,\forall \Lambda \in \s,\; \forall f \in \mathcal{F}_{\rm{loc}}
%\ee
%or, in an even shorter form, if and only if 
%\be \label{DLR3}
%\forall \Lambda \in \s,\; \mu \gamma_\Lambda = \mu
%\ee
We denote by $\mathcal{G}(\gamma)$ the set of measures consistent with $\gamma$. 
%For a translation-invariant specification, $\mathcal{G}_{\rm{inv}}(\gamma)$ is the set of translation-invariant elements of $\mathcal{G}(\gamma)$.
\end{definition}

%Different DLR-measures share the same expression for their conditional probabilities w.r.t. the outside of {\em finite} sets, thanks to the local specification (\ref{Dys}),  but the expressions are valid {\em almost surely} w.r.t to the DLR-measure itself. It is important to notice that this does not hold anymore when one wants to get the conditional probabilities w.r.t. the outside of {\em infinite} sets.
% for which highly phase-dependent asymptotic effects could yield different expressions depending on the DLR-measure considered. For this reason, 
The extension of the DLR equation to infinite sets is direct in case of uniqueness of the DLR-measure for a given  specification \cite{FP, Foll, Gold2}, but can be more problematic otherwise: it is valid for finite sets only and severe measurable problems can arise in case of phase transitions. Beyond the uniqueness case, such an extension was made possible by Fern\'andez and Pfister  \cite {FP} in the case of attractive models.
% and, as we will make essential use of it, we describe it now in our particular case. 
The terminology used  is that of  {\em global specifications}, and this is in fact a central tool in studying various  Gibbs vs. non-Gibbs questions.

%\subsection{Global specification}

\begin{definition}[Global specification \cite{FP}]\label{Glob}:\\
A {\em global specification} $\Gamma$ on $\Z$ is a family of probability kernels $\Gamma=(\Gamma_S)_{S \subset \Z}$ on $(\Omega_S,\mathcal{F}_{S^c})$ such that for {\em any} $S$ subset of $\Z$:
\begin{enumerate}
%\item $\Gamma_S(\cdot | \omega)$ is a probability measure on $(\Omega,\mathcal{F})$ for all $\omega \in \Omega$.
%\item $\Gamma_S(A | \cdot)$ is $\mathcal{F}_{S^c}$-measurable for all $A \in \mathcal{F}$.
\item $\Gamma_S(B|\omega)=\mathbf{1}_B(\omega)$ when $B \in \mathcal{F}_{S^c}$.
\item For all $S_1 \subset S_2 \subset \Z$, $\Gamma_{S_2} \Gamma_{S_1}=\Gamma_{S_2}$ where the product of kernels is made as in (\ref{DLR0}).
\end{enumerate}
%\end{definition}
%Similarly to the consistency with a (local) specification, one introduces the {\em compatibility of measures with a global specification}.
%\begin{definition}
%Let $\Gamma$ be a global specification. 
We write $\mu \in \mathcal{G}(\Gamma)$, or say that  $\mu \in \mathcal{M}_1^+$ is $\Gamma${\em -compatible}, if for all $A \in \mathcal{F}$ and {\em any} $S \subset \Z$,
\be \label{DLR4}
\mu[A|\mathcal{F}_{S^c}](\omega)=\Gamma_S(A|\omega), \; \mu{\rm -a.e.} \;  \omega.
\ee
\end{definition}

%Note, by considering $S=\mathbb{Z}$, that $\mathcal{G}(\Gamma)$ contains at most one element. \\

%***************************************\\
%Blabla or complete or indicate decimated...\\
%***************************************

\subsection{Gibbs and Quasilocal Measures}

{\bf A specification is said to be  quasilocal} when the set of quasilocal functions is conserved by its kernels. More formally, for any local function, its image by the kernels constituting $\gamma$ should be a continuous function of the boundary condition :
\be\label{qlocmes}
\gamma \; {\rm quasilocal} \; \; \Longleftrightarrow \; \; \gamma_\Lambda f \in C(\Omega) \; {\rm for \; any} \; f \; {\rm local} \;  ({\rm or \; any} \; f\;{\rm  in}\; C(\Omega)).
\ee
{\bf A measure is said to be quasilocal} when it is specified by a quasilocal specification. 

In fact, such quasilocal measures are very close to {\em Gibbs measures}, originally designed to represent equilibrium states satisfying a variational principle for a  (formal) Hamiltonian $H$. The latter is defined {\em via} a potential $\Phi$, i.e. a family $(\Phi_A)_{A \in \s}$ of local functions $\Phi_A \in \mathcal{F}_A$ that provide the contributions of spins in finite sets $A$ to the total energy through the {\em finite-volume Hamiltonians} -- or {\em Hamiltonians with free boundary conditions} -- defined for all $\Lambda \in \s$ by
\be \label{Ham}
H_\Lambda(\omega)=\sum_{A \subset \Lambda} \Phi_A(\omega),\; \forall \omega \in \Omega.
\ee
To define Gibbs measures, we require $\Phi$ to be  {\em Uniformly Absolutely Convergent} (UAC), i.e. that $\sum_{A \ni i} \sup_\omega |\Phi_A(\omega)| < \infty, \forall i \in \Z$.
%\be \label{UACPot}
%\forall i \in \Z,\;\sum_{A \ni i} \sup_\omega |\Phi_A(\omega)| < \infty.
%\ee
%For such a potential, 
One can give sense to the  {\em Hamiltonian at volume $\Lambda \in \s$ with boundary condition $\omega$} defined for all $\sigma \in \Om$ as $H_\Lambda^\Phi(\sigma | \omega) := \sum_{A \cap \Lambda \neq \emptyset} \Phi_A(\sigma_\Lambda \omega_{\Lambda^c}) (< \infty)$.
%\be \label{Hambc}
%H_\Lambda^\Phi(\sigma | \omega) := \sum_{A \cap \Lambda \neq \emptyset} \Phi_A(\sigma_\Lambda \omega_{\Lambda^c}) (< \infty).
%\ee

The {\em Gibbs specification at inverse temperature $\beta>0$} is then defined by
\be \label{Gibbspe}
\gamma_\Lambda^{\beta \Phi}(\sigma \mid \omega)=\frac{1}{Z^{\beta \Phi}_\Lambda(\omega)} \; e^{-\beta H_\Lambda^\Phi(\sigma | \omega)} (\rho_\Lambda\otimes \delta_{\omega_{\Lambda^c}}) (d \sigma)
\ee
where the normalization $Z_\Lambda^{\beta \Phi}(\omega)$ -- the partition function --is a normalizing constant related to free energy and pressure. Such a specification is {\em non-null}\footnote{In the sense that $\forall \Lambda \in \s,\; \forall A \in \mathcal{F}_\Lambda$, $\rho(A)>0$ implies that $\gamma_\Lambda (A | \omega) >0$ for any  $\omega \in \Om$.  This property sometimes is also called the ``finite-energy'' property.} and has the property that it is {\em quasilocal}, thanks to the  convergence properties  of the defining potential (see e.g. \cite{HOG,ALN2}). Gibbs measures are those consistent with a Gibbs specification defined in terms of a UAC potential, but Kozlov \cite{Ko} and Sullivan \cite{Su} established that   being Gibbs is in fact also equivalent to being non-null and quasilocal. We take then the following 
\begin{definition}[Gibbs measures]:\\
$\mu \in \mathcal{M}_1^+$ is a Gibbs measure iff $\mu \in \mathcal{G}(\gamma)$,  where $\gamma$ is a non-null and quasilocal specification.
\end{definition}
While non-nullness prevents  hard-core exclusions and  only allows  a proper exponential factor to alter the product structure of the measure -- to get correlated random fields --, quasilocality allows us to interpret Gibbs measures as natural extensions of the class of  Markov fields\footnote{In fact Sullivan used the term of {\em Almost Markovian} instead of quasilocal in \cite{Su}.}.\\

%**************************************************************************************\\
%Here, in the following 20 lines, I did not change anything for referee 1, I only try to reduce a bit to satisfy ref 2. Maybe this can be reduced again. I don't exactly understand the remark of referee 1, because we precise before the almost sure definition, and also afterwards. Need to think about it a bit more to find a good answer, but this is not a big problem\\
%**************************************************************************************\\

Indeed, when $\mu \in \mathcal{G}(\gamma)$ is quasilocal, then for any  $f$ local and  $\Lambda \in \s$,  the conditional  expectations of $f$ w.r.t. the outside of $\Lambda$ are $\mu$-a.s. given by $\gamma_\Lambda f$, by  (\ref{DLR0}), and this is itself a continuous function of the boundary condition by (\ref{qlocfu})  when the continuous version of the conditional probability, which exists,  is chosen. Thus,  for this version, one gets for any $\omega$
\be \label{esscont}
\lim_{\Delta \uparrow \mathbb{Z}} \sup_{\omega^1,\omega^2 \in \Omega}  \Big| \mu \big[f |\mathcal{F}_{\Lambda^c} \big](\omega_\Delta \omega^1_{\Delta^c}) - \mu \big[f |\mathcal{F}_{\Lambda^c} \big](\omega_\Delta\omega^2_{\Delta^c})\Big|=0
\ee
which yields an (almost-sure) asymptotically  weak dependence on the conditioning.
%,
%which can be seen as an extended  Markov property. 
In particular, for Gibbs measures the conditional probabilities always have continuous versions, or equivalently
%In particular, for Gibbs measures, it is not possible to change  its conditional probabilities on exceptional sets in order to get a discontinuous version: one says that 
 there is no point of essential discontinuity:
\begin{definition}[Essential discontinuity]\label{essdiscdef}:\\
A configuration $\omega$ is said to be {\em a point of essential  discontinuity} for a conditional probability of $\mu \in \mathcal{M}_1^+$ if  no version of the conditional probability is continuous at that point. Such a point is thus a point of discontinuity for each specification compatible with the prescribed conditional probabilities.
\end{definition}

To get such a "bad" configuration $\omega$, it is sufficient  that there exists $\Lambda_0 \in \s$, $f$ local, $\delta >0$, such that for all $\Lambda$ with $\Lambda_0 \subset \Lambda$ there exist  $\mathcal{N}_\Lambda^1(\omega)$ and $\mathcal{N}_\Lambda^2(\omega)$,   two open\footnote{ or at least positive-measure, compare \cite{Fer,EEIK}.} neighborhoods of $\omega$ 
%in which all configurations are equal to  $\omega$ inside $\Lambda$,
 on which  all versions the conditional expectations of $f$ differ substantially, by more than $\delta$.\\
 
  To be a bit more specific, there exists in this case even an everywhere discontinuous 
specification $\gamma$ : 
%one can always find one configuration, ( in fact many configurations) $\omega_1 \in  \mathcal{N}_\Lambda^1(\omega),\;$ and another set of configurations  $\omega_2 \in  \mathcal{N}_\Lambda^2(\omega),$ such that   
%$$
%\Big| \mu \big[f |\mathcal{F}_{\Lambda^c} \big](\omega^1) - \mu \big[f |\mathcal{F}_{\Lambda^c} \big](\omega^2)\Big| > \delta.
%$$
one can find a $\delta >0$ and  for any $n$ one can find  volumes $\Lambda_n$, increasing in $n$, and $V_n$ much larger than and  dependent on $\Lambda_n$, such that for all $\omega^i \in \mathcal{N}_{\Lambda}^i(\omega)$, $i=1,2$ and all $\sigma'$, 
$$
 \big| \gamma(f| \omega_{\Lambda_n} \omega^1_{{V_n \backslash \Lambda_n}} \sigma'_{V_{n}^{c}}) - 
\gamma(f| \omega_{\Lambda_n}\omega^{2}_{V_n \backslash \Lambda_n} \sigma'_{V_{n}^{c}}) \big| > \delta.
$$ 
Then any other specification with the same conditional probabilities is necessarily also discontinuous. (One can change the above expression only for a measure-zero set of $\sigma'$).\\

Equivalently, one gets in integrated form: 
For a local function $f$, ${\mu}_{\Lambda_0}[f|\cdot]$ is $\mu$-essentially discontinuous at $\omega$, if there exists an $\varepsilon>0$ such that 
 \begin{equation}\label{essdisc}
\displaystyle\limsup_{\Lambda\uparrow\infty}\sup_{\xi^1,\xi^2\atop{\Lambda'\supset\Lambda \atop{|\Lambda'|<\infty}}}\vert {\mu}_{\Lambda_0}[f|\omega_{\Lambda\setminus\Lambda_0} \xi^1_{\Lambda'\setminus\Lambda}]-{\mu}_{\Lambda_0}[f|\omega_{\Lambda\setminus\Lambda_0} \xi^2_{\Lambda'\setminus\Lambda}]\vert > \varepsilon  .
\end{equation}
%\end{definition}

%\be\label{essdisc}
%\lim_{\Delta \uparrow \mathbb{Z}} \sup_{\omega^1,\omega^2 \in \Omega}  \Big| \mu \big[f |\mathcal{F}_{\Lambda^c} \big](\omega_\Delta \omega^1_{\Delta^c}) - \mu \big[f |\mathcal{F}_{\Lambda^c} \big](\omega_\Delta\omega^2_{\Delta^c})\Big|  > \delta.
%\ee
%\end{definition}
In the generalized Gibbsian framework, one also says that such a configuration is a {\em bad configuration} for the considered measure, see e.g. \cite{ALN2}.
The existence of such bad configurations implies non-Gibbsianness of the associated measures.

\subsection{Dyson-Ising models: Ferromagnets in One Dimension}

In our  framework\footnote{Or more generally when the configuration space is {\em standard Borel}, see \cite{HOG}.}, for any given $\mu \in \mathcal{M}_1^+$, it is always possible to construct a specification $\gamma$ such that $\mu \in \mathcal{G}(\gamma)$ (see e.g.  Goldstein \cite{Gold}, Preston \cite{Pr} or Sokal \cite{Sok}). Nevertheless, even in such a framework, there exist specifications $\gamma$ for which $\mathcal{G}(\gamma)=\emptyset$ (see e.g. \cite{HOG, ALN2}), others where $\mathcal{G}(\gamma)=\{\mu\}$ but also -- and this is more interesting for us -- some for which this set contains more than one element. In the latter, we say in mathematical statistical mechanics that there is a {\em phase transition}. The set of DLR measures is then known to be a convex set whose extremal elements are trivial on the tail $\sigma$-algebra $\mathcal{F}_\infty$. Any other element of $\mathcal{G}(\gamma)$ admits a unique\footnote{It is a {\em Choquet simplex}, see \cite{Dy, HOG}.} convex combination of the extremal elements and is characterized by its action on the tail $\sigma$-algebra $\mathcal{F}_\infty$ \cite{VEFS, HOG}. We focus here on such a case in dimension one:

\begin{definition}[Dyson-Ising model]:\\
Let $\beta >0$ be the inverse temperature  and consider $1 < \alpha \leq 2$. We call {\em Dyson-Ising specification} with decay parameter $\alpha$ the Gibbs specification (\ref{Gibbspe}) with (pair-)potential $\Phi^D$ defined  for all $\omega \in \Omega$ by

\be \label{Dys}
\Phi_A^D(\omega)= - \frac{1}{|i-j|^\alpha} \om_i \om_j \; {\rm when} \; A=\{i,j\} \; \subset \mathbb{Z},\; {\rm and}\; \Phi_A^D \equiv 0 \; {\rm otherwise}.
\ee

%\be \label{LRDysonSpe}
%\gamma_\La^D(d \si | \om) = \frac{1}{Z_\La^\beta(\om)} \; e^{\beta \sum_{i \neq j, i \in \La, j \in \Z} \frac{1}{|i-j|^\alpha} \si_i \si_j} \; \rho_\La \otimes \delta_{\om_{\La^c}} (d \si)
%\ee
%where the normalization $Z_\La^\beta(\om)$ is the usual partition function.

We shall also need to consider Dyson models with non-zero magnetic field $h \in \mathbb{R}^*$ for which one also has a self-interaction part
$
\Phi_A^D(\omega)= - h \omega_i  \; {\rm when} \; A=\{i\}  \; \subset \mathbb{Z}
$
\end{definition}

The Dyson-Ising specification is {\em monotonicity-preserving} 
(or {\em attractive}) in the sense that for all bounded increasing functions 
$f$, and  $\La \in \s$, the function $\gamma_\La^D f$ is 
increasing.\footnote{It a consequence of the FKG
property \cite{FKG,Hul}:  spins have a tendency to align.} 
%balanced by entropic effects.}. 
Using as boundary conditions the extremal (maximal and minimal) elements of this
order $\leq$ already allows to define the extremal elements of $\mathcal{G}(\gamma^D)$. Indeed, one can learn in e.g. \cite{FP,Hul,Leb} that
\begin{proposition}\label{Wlimit}:\\
The weak limits
\be \label{muplusminus}
\mu^-(\cdot) := \lim_{\La \uparrow \mathbb{Z}} \gamma_\La^D (\cdot | -)\; \; {\rm and} \; \; \mu^+(\cdot) := \lim_{\La\uparrow \mathbb{Z}}  \gamma_\La^D (\cdot | +)
\ee
are well-defined, translation-invariant and extremal elements of $\mathcal{G}(\gamma^D)$. For any $f$ bounded increasing, any other measure $\mu \in \mathcal{G}(\gamma^D)$ satisfies
\be \label{stochdom}
\mu^-[f] \leq \mu[f] \leq \mu^+[f].
\ee
Moreover, $\mu^-$ and $\mu^+$  are respectively left-continuous and right-continuous.
\end{proposition}

When the range is long enough ($1<\alpha \leq 2$), it is possible to recover in dimension one low-temperature behaviours usually 
%devoted to 
 associated to
higher dimensions for the standard Ising model, and we quote here those used in this paper.

\begin{proposition}\label{DyFrSp}:\\
\begin{enumerate}
\item The Dyson-Ising model with potential (\ref{Dys}), for $1< \alpha \leq 2$, exhibits a {\em phase transition at low temperature}:
$$
\exists \beta_c^D >0, \; {\rm such \; that} \; \beta > \beta_c^D \; \Longrightarrow \; \mu^- \neq \mu^+ \; {\rm and} \; \mathcal{G}(\gamma^D)=[\mu^-,\mu^+]
$$
where the extremal  measures $\mu^+$ and $\mu^-$ are translation-invariant\footnote{Furthermore, all Gibbs measures for our Dyson-Ising models are translation-invariant  (\cite{HOG}, Theorem 9.5).}. They have in particular opposite magnetisations   $\mu^+[\sigma_{0}]=-\mu^-[\sigma_{0}]=M_0(\beta, \alpha)>0$ at low temperature.
\item {\em Uniqueness in non-zero magnetic field} : The Dyson Ising model in a homogeneous field $h$ has a unique Gibbs measure. 
\end{enumerate}
\end{proposition}

{\bf Proofs:} \\
The existence of phase transitions at low temperature comes was first proved by  Dyson for $1<\alpha<2$ \cite{Dys} and  Fr\"ohlich/Spencer for $\alpha=2$ \cite{frsP}. \\
Uniqueness in non-zero field follows immediately from a theorem given in the Appendix of \cite{Rue72} which applies to all ferromagnetic Ising pair interactions, including Dyson models. The proof  uses the Lee-Yang  circle theorem to obtain an analyticity property of the pressure, as well as the FKG stochastic domination. See also \cite{HOG}, Notes to Chapter 16.2,  or the detailed proof of \cite{FV} in the standard Ising case.

%************************************************\\
%{\em Yang-Lee etc.To be written, if you agree to put it here. MAy be quote the explanation of \cite{FV} what zeroes of partition have to do with uniqueness via analycity of the pressure, to answer ref 1.}\\
%******************************************\\
%\item Equivalence with $+$ or $-$ b.c. for non negative (resp. non positive) $h>0$ (resp. $h<0$).
%****************{\em same remark. I took  the statement of \cite{Leb2} but i should read and write again on that, i think this could help to explain/describe the prrof. Again, it might be not necessary...}**************\\

%{\bf ?? I think the uniqueness statement above is enough??}

{\bf Remark 1:}\\ 
The infinite-volume limit of a state in which there is a $+$ (resp. $-$)-measure or a Dyson model in a field $h >0$ (resp. $h<0$) outside 
%the boundaries
  is the same  $+M_0(\alpha,\beta)$  (resp. $-M_0(\alpha, \beta)$) as that obtained from $+$ (resp. $-$)-boundary conditions (independent of the magnitude of $h$).  This can be e.g. seen by an extension of the arguments of \cite{LebP}, see also \cite{Leb2}. Notice that taking the $+$-measure of the zero-field Dyson model outside a finite volume enforces this same measure inside (even before taking the limit); adding a field makes it it more positive, and taking the thermodynamic limit then recovers the same measure again.

To express the conditional magnetisations of the decimated measures on different sub-neighborhoods of the  alternating configuration, we need to extend the (local) Dyson-Ising specification into a global one, in the low-temperature phase transition region. Note that both the decimated lattice and its complement are infinite, which is why the existence of a global specification is very convenient.
Following the construction of  \cite{FP} in the general monotonicity-preserving case, we get:
\begin{theorem}\label{globspe}:\\
Consider any Dyson-Ising model on $\Z$ at inverse temperature $\beta >0$, i.e. the specification $\gamma^D$ with potential (\ref{Dys}) and its extremal Gibbs measures $\mu^+$ and $\mu^-$ defined by (\ref{muplusminus}). Define $\Gamma^+=\big(\Gamma_S^+\big)_{S \subset \Z}$ to be the family of probability kernels on $(\Omega, \mathcal{F})$  as follows:
\begin{itemize}
\item For $S=\Lambda$ finite, for all $\omega \in \Omega$, $\Gamma^+_\Lambda(d \sigma | \omega) := \gamma^D_\Lambda (d \sigma |  \omega).$
\item For $S$ infinite, for all $\omega \in \Omega$,
\be  \label{Globalmu+}
\Gamma^+_S(d\sigma | \omega):=\mu_S^{+,\omega} \otimes \delta_{\omega_{S^c}}(d \sigma)
\ee
where $\mu_S^{+,\omega}$ is the constrained measure on $(\Omega_S,\mathcal{F}_S)$  (well-)defined as the weak limit 
\be \label{const}
\mu_S^{+,\omega}(d \sigma_S):=\lim_{I \in \s, I \uparrow S} \gamma^D_I (d \sigma\mid +_S \omega_{S^c}).
\ee

%with more explicitly 
%$$
%\gamma^I_\Delta (d \sigma | +_S \omega_{S^c}})= f(\sigma_\Delta +_{S \setminus \Delta} \omega_{\Delta^c}) \rho_\Lambda (d \sigma_\Delta) %\otimes \delta_{+_{S %\setminus \Delta}}(d +_{S \setminus \Delta}) \otimes \delta_{\omega_{S^c}})(d \sigma_{\Delta^c})}{Z_\Lambda(+_S \omega_{S^c})}}$$
%and
%$$
%f(\sigma_\Delta +_{S \setminus \Delta} \omega_{\Delta^c})= \frac{1}{Z_\Delta(\omega)} \; e^{\beta(\sum_{\langle ij \rangle \subset \Delta} \sigma_i \sigma_j + %\sum_{\langle ij \rangle,i \in \Delta, j \in S^c} \sigma_i \omega_j + \sum_{\langle ij \rangle,i \in S \setminus \Delta, j \in T^c} \omega_j + \sum_{\langle ij %\rangle,i \in \Delta, j \in S \setminus \Delta} \sigma_i }). 
%$$
\end{itemize}
Then $\Gamma^+$ is a global specification such that $\mu^+ \in \mathcal{G}(\Gamma^+)$. It is moreover monotonicity-preserving and right-continuous. Similarly, one defines a monotonicity-preserving and left-continuous global specification $\Gamma^-$ such that $\mu^- \in \mathcal{G}(\Gamma^-)$.
\end{theorem}

Remark that when the set $S$ is infinite, one proceeds in two steps,  the  order of which is crucial: Freeze first the configuration into $\omega$ on $S^c$ and  perform afterwards the weak limit with $+$-boundary condition {\em in} $S$, to get the constrained measure $\mu^{+,\omega}_S$ on $(\Omega_S,\mathcal{F}_S)$. Note also that    the global specification obtained need not to be quasilocal in general, even when the original specification is itself quasilocal. This failure of quasilocality, caused by  long-range ordering due to hidden phase transitions, is in fact crucial, as we see now.

\section{Decimation of the Dyson Ising Model}
\subsection{Set-up : Decimation Transformation}

We start at low temperature in the phase transition region of the Dyson-Ising model with any Gibbs measure $\mu$, mainly considering the $+$-measure $\mu^+$, obtained  as the weak limit (\ref{muplusminus}) with $+$-boundary conditions, and introduce the following {\em decimation transformation}:
\be \label{DefDec}
 T \colon (\Omega,\mathcal{F})  \longrightarrow (\Omega',\mathcal{F}')=(\Omega,\mathcal{F}); \; 
\omega \; \;   \longmapsto \omega'=(\omega'_i)_{i \in
\mathbb{Z}}, \; {\rm with} \;  \omega'_{i}=\omega_{2i}
\ee
This transformation acts on measures in a canonical way and we denote $\nu^+:=T \mu^+$ the decimation of the $+$-measure. It is formally defined as an image measure via
$$
\forall A' \in \mathcal{F'},\; \nu^+(A')=\mu^+(T^{-1} A')=\mu^+(A) \; {\rm where} \; A=T^{-1} A'= \big\{\omega: \omega'=T (\omega) \in A' \big\}.
$$
When necessary, we distinguish between original and image sets using  primed notation\footnote{Notice that by  rescaling  the configuration spaces $\Omega$ (original) and $\Omega'$ (image) are identical.}. 

%In particular, we get the following expressions in terms of global specifications first, expressed themselves in terms of  constrained measures afterwards, of conditional expectations of the spin at the origin, used next section to prove an essential discontinuity. \\

%****************************************************************\\
%No changes for the moment for the following paragraph. Maye we should check at the end that things are not told too many times in different forms...\\
%***********************************************************\\

We want to study the continuity of various conditional expectations
%, under $\nu^+$, 
under decimated Dyson measures
of the spin at the origin when the outside is fixed 
%under 
 in
some special configuration that we denote\footnote{It will be used for an alternating configuration in the proof, but here we do not use its particular form.}  $\omega'_{\rm alt}$. First note that
%As the conditioning takes place on an infinite set with infinite complement, we need here  global specifications for the decimated measures built as in Theorem \ref{globspe} got from  \cite{FP}.
% or \cite{ALN}. To build these specifications, we first note that
%On the other hand, if in the "annulus" we impose all plus spins for the primed sites, our magnetisation at the origin will be larger than that of the plus state of the constrained system.  In fact, due to the fact that the Dyson model in a positive external field has only one Gibbs measure, the influence transmitted through the annulus decays with the distance to the external boundary, whatever the boundary condition. A similar argument applies to the situation in which the primed spins in the annulus are all minus.  
%{\bf REMARK, the nonGibbsianness can be proven at temepratures strictly below the phase transition temperature of %the Dyson models (the constrained model has a lower transition temperature)}
%To do so, consider a basis of neighborhood $(\mathcal{N}_{\Lambda'}(\omega'_{\rm alt})_{\Lambda' \in \mathcal{S}}$  %for $\Lambda' \in \s$ containing the origin, and any $\omega' \in \mathcal{N}_{\Lambda'}(\omega'_{\rm alt})$, we %express the conditional expectations  using the very definition of $\nu^+$ as an image measure of $\mu^+$ via the %decimation transformation $T$: for any  $\omega' \in \mathcal{N}_{\Lambda'}(\omega'_{\rm alt})$ and any  $\omega \in %T^{-1} \{\omega'\}$,
\be \label{condmagn}
 \nu^+[\sigma'_0 | \mathcal{F}_{\{0\}^c} ](\omega') = \mu^+[\sigma_0 | \mathcal{F}_{S^c} ](\omega),\; \nu^+{{\rm -a.s.}}
\ee
where $S^c=(2 \mathbb{Z}) \cap \{0\}^c$, i.e. with $S= (2 \mathbb{Z})^c \cup \{0\}$ is not  finite: {\em the conditioning  is  not on the complement of a finite set}. We need thus to use the global specification $\Gamma^+$ such that $\mu^+ \in \mathcal{G}(\Gamma^+)$, built in Theorem \ref{globspe}, with $S= (2 \mathbb{Z})^c \cup \{0\}$ consisting of the {\em odd integers plus the origin}.  Hence $S =(2 \mathbb{Z})^c \cup \{0\}$ and (\ref{condmagn}) yields for 
%$\nu^+$-a.e.
 all, (using the specification property)
 $\omega' \in \mathcal{N}_{\Lambda'}(\omega'_{\rm alt}) $ and $\omega \in T^{-1} \{\omega'\}$: 
\be \label{condmagn2}
 \nu^+[\sigma'_0 | \mathcal{F}_{\{0\}^c} ](\omega') = \Gamma_{S}^+ [\sigma_0 | \omega] \; \; \mu^+{\rm -a.e.} (\omega). 
\ee

%({\bf that will be defined later as the alternating configuration, so that $\omega_{2i}=(-1)^i$)}.
 Now, by (\ref{Globalmu+}) we have an expression of the latter in terms of the constrained measure $\mu^{+,\omega}_{(2\mathbb{Z})^c \cup \{0\}}$, with $\omega \in T^{-1} \{\omega'\}$  so that we get for any 
%{\bf ( ? $\nu^+$-a.e ?)} 
$\omega' \in \mathcal{N}_{\Lambda'}(\omega'_{\rm alt})$,
$$
\nu^+[\sigma'_0 | \mathcal{F}_{\{0\}^c} ](\omega') = \mu^{+,\omega}_{(2\mathbb{Z})^c \cup \{0\}} \otimes \delta_{\omega_{2\mathbb{Z} \cap \{0\}^c}} [\sigma_0].
$$
Thanks to monotonicity-preservation, the constrained measure is explicitly built as the weak limit  (\ref{const}) obtained by $+$-boundary conditions fixed after a freezing 
%{\bf in} 
$\omega$ on the even sites :
\be \label{constrLimit}
\forall \omega' \in \mathcal{N}_{\Lambda'}(\omega'_{\rm alt}), \forall \omega \in T^{-1} \{\omega'\},\;   \mu^{+,\omega}_{(2\mathbb{Z})^c \cup \{0\}} (\cdot) =\lim_{I \in\s,I \uparrow (2 \mathbb{Z})^c  \cup \{0\}} \gamma^D_I (\cdot\mid +_{(2 \mathbb{Z})^c  \cup \{0\})} \omega_{2 \mathbb{Z} \cap\{0\}^c}).
\ee
and  it is enough to consider this limit on a sequence of intervals  $I_n=[-n,+n] \cap \mathbb{Z}$ in the original space.
Now, one  obtains an essential discontinuity if we can get an  difference in the expectation of the spin at the origin of this constrained measure conditioned on two different open subsets  of arbitrary neighborhoods of $\omega'_{\rm alt}$. As we shall see, this is indeed the case as soon as the temperature is low enough in order to get a phase transition for the Dyson-Ising ferromagnet on the odd sites -- the hidden phase transition --. \\

This type of transformation was also the basic example in 
%of the seminal work of van Enter {\em et al.}
 \cite{VEFS}, where non-quasilocality is proved in dimension 2 at low enough temperature, as soon as a phase transition is possible for an Ising model  on the decorated lattice, which consists of a version of $\mathbb{Z}^2$ where the "even" sites have been removed. In our one-dimensional set-up, the role of this decorated lattice will be 
%equivalent {\em to} %of 
played by the set of odd sites, $2 \mathbb{Z}+1$, which again can be identified with 
$\mathbb{Z}$ itself. We observe that when a  phase transition holds for the Dyson specification -- at low enough temperature for $1 < \alpha \leq 2$ -- the same is true for the constrained specification (\ref{const}) {\em with alternating constraint}, albeit one needs even lower temperatures to have a phase transition. This leads   to non-Gibbsianness of $\nu^+$. Once the $+$-measure is shown to be non-Gibbsian after being subjected to a decimation transformation, the same holds true for all other Gibbs measures of the model.   
% We shall come back to this later, before we state and prove our main result. 

\subsection{Non-Gibbsianness at Low Temperature}
\begin{theorem}\label{thm2}:\\
For any  $1<\alpha \leq 2$, at low enough temperature
% $\beta > \beta_c^D$, 
the decimation  $\nu$ of any 
%$+$-phase
 Gibbs measure $\mu$ of the Dyson-Ising model, $\nu=T \mu$ is non-quasilocal, hence non-Gibbs.
\end{theorem}

{\bf Sketch of Proof:}\\ We know from Section 2.2 -- and basically from \cite{VEFS} -- that to get non-Gibbsianness, it suffices to find  an essential discontinuity, i.e.  a local function $f$, a finite subset $\Lambda'$ and a configuration $\omega'$ so that the conditional expectation of $f$ when $\Lambda'^c$ is fixed under $\omega'$ cannot be made continuous by changes on zero-measure sets.
%, i.e. by taking other versions. 
Such a point of essential discontinuity is also called a  {\em bad configuration}. Here, the bad configuration for %the image measure 
$\nu^+$ will be, just as  in \cite{VEFS} in the two-dimensional case, the so called {\em alternating configuration} $\omega'_{{\rm alt}}$ defined for any $i \in \mathbb{Z}$ as $(\omega'_{\rm alt})_i=(-1)^i$. To get the essential discontinuity, the choice of $f(\sigma')=\sigma'_0$ will be enough.\\

%***********************************************************\\
%No change or this paragraph, except quoting propoition 2, same remark as above...\\
%***********************************************************\\
% Intuitively, 
{\bf Observation}:\\
Because any non-fixed site at all odd distances has a positive and a negative spin whose influences cancel, conditioning by this alternating configuration yields a constrained model that is again a model of Dyson-type.  Indeed, it is a Dyson model at zero field at a temperature which is higher by $2^{\alpha}$, which again has a low-temperature transition in our range of decays $1 < \alpha \leq 2$.
 The coupling constants are multiplied by a factor $ 2^{- \alpha}$, due to only even distances occurring. Thus the argument will only work if the temperature is at least  smaller by that factor than the transition temperature of the original Dyson model. \\

The non-Gibbsianness proof essentially goes along the lines sketched in \cite{VEFS}, with the role of the ``annulus'' played by two large intervals $[-N,-L-1]$ and $[L+1,N]$ (with $N$ {\em much} larger than $L$) to the left and to the right of the central interval $[-L,+L]$. If we constrain the spins in these two intervals to be either  $+$ or $-$, within these two intervals the measures on the unfixed spins are close to those of the Dyson-type model in a positive, c.q. negative, magnetic field. As those measures are unique (due to
% e.g. 
 FKG and a Yang-Lee argument \cite{YL},  as discussed  in Proposition \ref{DyFrSp}, see  also \cite{Ker}), no influence from the boundary can be transmitted 
%by 
 via 
this ``annulus''.\\

 Due to the long range of the Dyson interaction, there may be also a direct influence from the boundary, that is from  beyond the annulus, to the central interval, however. But by choosing $N(L)$ large enough -- e.g.  $N =L^{\frac{1}{\alpha-1}}$  -- we can make this direct  influence as small as we want, so the strategy of \cite{VEFS}, there worked out for finite-range models,  does also work here. The special configuration chosen is  also an  alternating one (just as in \cite{VEFS}). 
Conditioned on all primed spins being alternating, the conditioned model is a Dyson-like model in zero field, due to cancellations, so that a phase transition occurs at low temperature, making it possible to select the phase by boundary conditions arbitrarily far away.  On the contrary, when conditioned on all primed spins to be $+$ (resp. $-$), there is no phase transition, but the system of unprimed spins 
has a unique Gibbs measure.  
It is a Dyson model, again at a heightened temperature, but now in a homogeneous external field,   with positive (resp. negative) magnetisation $+M_0(\beta,\alpha)>0$ (resp. $-M_0(\beta,\alpha)<0$), stochastically larger  (resp. smaller) than the zero-field $+$ ( resp. $-$)-measure. What thus has to be shown is that it is possible to prescribe $+$ or $-$ spins on a large enough annulus so that they select the above  measures, which then can act similar to ``pure'' boundary conditions,   whatever is put outside, on the boundary beyond the annulus.%\\

\begin{lemma}\label{keylemma}:\\
%Let $\beta$ > \beta^D_c$ and c
Consider a Dyson-Ising model with 
%long-range 
decay parameter  $1 < \alpha \leq 2$, at sufficiently low temperature. Let $\Lambda' \subset \Delta' \in \mathcal{S}$ and consider two arbitrary configurations $\omega'^+ \in \mathcal{N}_{\Lambda',\Delta'}^+(\omega'_{\rm alt})$ and $\omega'^- \in \mathcal{N}_{\Lambda',\Delta'}^-(\omega'_{\rm alt}) $. Then  $\exists \delta >0$, and  $\exists \Lambda'_0$ big enough s.t. for some $\Delta' \supset \Lambda' \supset \Lambda'_0$ with $\Delta' \setminus \Lambda'$ chosen big enough compared to $\Lambda'$, 
%such that, 
for all $\omega^+ \in T^{-1} \{\omega'^+\}$ and all $\omega^- \in T^{-1} \{\omega'^-\} $ 
\be \label{keymagn}
\Big| \mu^{+,\omega^+}_{(2\mathbb{Z})^c \cup \{0\}}[\sigma_0] -  \mu^{+,\omega^-}_{(2\mathbb{Z})^c \cup \{0\}}[\sigma_0] \Big| > \delta.
\ee
\end{lemma}

{\bf Proof of Lemma \ref{keylemma}:}\\
Let us  first choose the annulus large enough that we can neglect boundary effects beyond {\bf $\Delta'$}, i.e. large  enough  that local  expectations are almost insensitive to  boundary effects, when the annulus increases properly. With the notation of the lemma, denote
$$
M^+ = \mu^{+,\omega^+}_{(2\mathbb{Z})^c \cup \{0\}}[\sigma_0] \; {\rm and} \; M^- = \mu^{+,\omega^-}_{(2\mathbb{Z})^c \cup \{0\}}[\sigma_0].
$$

%They are both magnetisations for some $+$-phase, but for different constrained specifications (....!!!!!!!!!!!!*********).

 % Afterwards we shall check that changes inside the annulus  will on the contrary substantially change local expectations.\\

Write $\Lambda'=\Lambda'(L)=[-L,+L]$ and $\Delta'=\Delta'(N)=[-N,+N]$, with $N >L$ and denote formally  by $H$ the Hamiltonian of both constrained specifications. We prove here that one can bound uniformly in $L$ the relative Hamiltonians with either $\omega_1^+$ and $\omega_2^+$ b.c. to get
\be\label{bc}
\Big| H_{\Lambda,\omega_1^+}(\sigma_\Lambda) - H_{\Lambda,\omega_2^+}(\sigma_\Lambda) \Big| \leq C <  \infty.
\ee
as soon as one takes $N=N(L)=O(L^{\frac{1}{\alpha -1}})$. Then one gets by \cite{BLP} (see also \cite{FV}) that all of the limiting Gibbs states obtained by these boundary conditions have an equivalent decomposition into extremal Gibbs states\footnote{Presumably trivial here, as the Gibbs measure will be unique, as we shall see.}  with the same measure zero sets, and thus yield the same magnetisation : $M^+=M^+(\omega, N, L)= M^+(\omega_1^+, N, L)=M^+(\omega_2^+, N, L)$ is indeed independent of $\omega$ as soon as it belongs to the pre-image of the $+$-neighborhood of the alternating configuration.\\

 To get (\ref{bc}), we use the long-range structure of the interaction to get a uniform bound
$$
\Big| H_{\Lambda,\omega_1^+}(\sigma_\Lambda) - H_{\Lambda,\omega_2^+}(\sigma_\Lambda) \Big| \leq 2 \sum_{x=-L}^L \sum_{k > N} \frac{1}{k^\alpha} < 2 L \frac{N^{1-\alpha}}{1-\alpha}
$$
so that $N=N(L)$ with $2 L \frac{N^{1-\alpha}}{\alpha - 1}=1$, or any bigger values of $N$, will do the
job.  So choose
\be \label{N}
N(L)=L^{\frac{1}{\alpha-1}}.
\ee
For example, for $\alpha=\frac{3}{2}$, one has thus to take some annulus of the order at least $N(L)=O(L^2)$.\\

Once we got rid of any possible direct asymptotic effects due to the long range, by choosing a large enough annulus as above, we now  check that changes inside the annulus will on the contrary substantially change local expectations $M^-$ or $M^+$ in the central interval.  These configurations  are drawn from neighborhoods of  the same  alternating configuration (which is still fixed inside the central interval).
%*********************************************\\
%I precise that its important to have alternate inside, to answer ref 2, mainly in a footnote for the moment. I di dnot change much, maybe come to this later.\\
%*********************************************\\
The main point is that 
%the frozen condition can 
freezing the primed spins to be  "$-$"  in a large enough annulus ( i.e. under the constraint $\omega^-$) can overcome 
%beat the weak limit with
the influence from the $+$-boundary condition  outside the annulus \footnote{From the initial measure, we decimate the $+$-state and this is visible in the weak limit with $+$-b.c. performed to get the global specification consistent with the decimated measure $\nu^+$.} when the frozen annulus $\Delta' \setminus \Lambda'$ is in a $-$-state AND the region around the origin is  frozen in an alternating configuration, for $L$ (and $N(L)$) large enough. 
%Thus the second expectation of (\ref{keymagn}) can be made as close as possible to 
 In the annulus the magnetisation of the -even-distance- Dyson-Ising model is essentially that of the model with a negative  homogeneous external field $-h$ everywhere, which at low enough temperature and for $L$ large enough is close to (in fact smaller than) the magnetisation of the Dyson-Ising model under the $-$-measure, i.e to $-M_0(\beta,\alpha) <0$ (and this $-$-measure is also unique, see \cite{Ker}).  Thus the inner interval where the constraint is alternating feels a $-$-like condition from outside its boundary. On the other hand, the magnetisation with the constraint $\omega^+$ will be close to or bigger 
%({\bf keep ?})
than $+M_0(\beta,\alpha)>0$ so that a  non-zero difference is created at low enough temperature. One needs  again to adjust the sizes of $L$ and $N$ to be sure that  boundary effects  from outside the annulus  are negligible  in the inner interval.\\
 
  Let us be a bit more precise now. We use the expression $(\ref{constrLimit})$ with $\omega'^+ \in \mathcal{N}^+_\Lambda(\omega'_{\rm alt})$ and to facilitate the proof we will make use of (\ref{constrLimit}), and freely change  between regular versions of conditional probabilities on 
%a thin, thinner than any other, 
arbitrarily small
neighborhoods of configurations (all $+$, all $-$, all $\omega'_{\rm alt}$, all $\omega^+$, etc.) with conditioning by the considered configuration itself (to avoid the problem of conditioning on zero measure sets). Recall that $\omega'^+$ is  generic for a configuration coinciding  with the alternating configuration around the origin, and with the  "$+$" one on the annulus depending on $N$ and $L$. To be still able to neglect boundary effects, we take $N(L)$ big compared to $L$ just as in the previous part of the proof.  Then we consider the homogeneous cases, all $+$ (resp. all $-$), that yields Dyson models with non-zero positive (resp. negative) homogeneous field), and to conclude we take $L$ (and hence $N(L)$) large enough to consider the $\omega^+$ (resp. $\omega^-$) as a small perturbation of it.  \\

%*********************
%Both $M^+$ and $M^-$ are magnetisation under {\em some} $+$ phases with constraint, one being got from the other by flipping to $+$ to $-$ in a large annulus, but the situation is not fully symmetric because we decimate the $+$-phase in the phase transition region. The more delicate case is probably the $-$ ones : who wins between the $-$'s in the annulus and the $+$-b.c. that has been used in the weak limit ? So we start to prove that there it is indeed possible to get a magnetisation under some minus phase, this fact being the main reason of our essential discontinuity.
%*********************

Conditioning of the primed sites to be all
%\footnote{Including inside $\Lambda$, it is not our magnetisation but we shall be close to this....MAYBE PRECISE IN THE TEXT} 
$+$ reduces $(\ref{constrLimit})$ to the magnetisation  obtained by taking a weak limit of a Dyson Ising specification with an everywhere\footnote{Modulo an adaptation to fix and unfix the spin at the origin, as in \cite{VEFS}.} homogeneous strictly positive external field.  This magnetic field is finite for $1<\alpha \leq 2$ and in our case the effect is even smaller because the $+$-b.c. is not present inside $\Lambda$ (but replaced by alternate spins whose effects cancel), so that one can take in this homogeneous case the non-zero magnetic field

%\footnote{Its value could probably be expressed  in terms of a Riemann $\zeta$-function.} 
$$
h^+ = 2 \sum_{k=L}^{+\infty} \frac{1}{(2k+1)^\alpha} := F(\alpha)  < \infty.
$$
 Thus, for the  naive choice of $\omega'=+$, the constrained magnetisation (the lhs of (\ref{condmagn})) is $+M_0({\beta, \alpha})$ of Proposition \ref{DyFrSp}, strictly positive at low temperature in our range $1<\alpha \leq 2$. \\

 Now, consider the case of $\omega'=\omega'^+$,  i.e. work on a neighborhood of $\omega'_{{\rm alt}}$ with an annulus filled with $+$. It reduces  again to a Dyson-Ising model with external field, but the latter $\big(h_x\big)_{x \in \mathbb{Z}}$ depends on $x \in \mathbb{Z}^d$ and  is not homogeneous anymore. Nevertheless, we observe that the difference with the homogeneous part is negligible 
%as the large cube
on most of the large "annulus intervals" 
 $I$ of (\ref{constrLimit}), and  the field is always non-negative, whether in the annulus or in the central interval.

 Indeed, in the annulus each site feels a strong positive field from all the $+$-constrained spins in the annulus, which dominates a possibly non-positive field due to either  the spins  outside or from the central interval. In the central interval, however, the spins just feel a $+$-field from the annulus, which will be weak when the distance from the site of the spin to the annulus increases, but still dominates the effects from the outside. The effect from the $-$ spins inside the interval is  canceled, either due to the positive spins from the alternating configuration in the central interval, or due to positive spins in the annulus.  
% Indeed, one can write, for all $x \in \mathbb{Z}$,
%$h_x=h^+ - \epsilon_x$
%where the perturbation $\epsilon_x$ depends on the location of $x$  as follows:

%\begin{eqnarray*}
%|x|> L : \epsilon_x= \sum_{k=-L}^L \frac{2}{(2k+1)^\alpha} = \circ(h^+) \; {\rm as} \; L \; \to \infty.\\
%-L \leq x \leq L : \epsilon_x \leq \sum_{k=-L}^L \frac{2}{(2k+1)^\alpha} = \circ(h^+) \; {\rm as} \; L \; \to \infty.\\
%\end{eqnarray*}

%{\bf This perturbation}  is  maximal for (the worst case) $x=0$ with a maximal perturbation dominated by the %homogenous one so that $h_0$ {\bf is of the order of} $h_+$ as $L$ goes to infinity.\\
%************************\\
%({\bf PRECISE SYMBOLS ? I eventualy don't know if it's necessary to write it like this.}). 
%*************************\\
More quantitatively, inside the central interval, when $|x|<L$, the field is larger than $O( L^{1- \alpha}) -O( N^{1- \alpha})$, which is small but positive, going to zero when $L$ and $N$ diverge. Inside the annulus, 
when $L< |x|< N$ the magnetic field is everywhere larger than $ \beta \big(1  - O(N^{1-\alpha})\big)$ which is strictly positive and uniformly lower-bounded. Deep inside the annulus the field approaches the homogeneous value, but the above observation already is enough for our proof.
%({\bf $beta$ ?. I need to check again}), strictly positive and uniformly. 
%{\bf Thus, using e.g. derivative of the pressure to get our constrained magnetisations, one get either negligible perturbations of magnetisations of order "plus" this of the $+$-measure, either "minus", }

A similar computation holds with the all $-$'s-constrained specification.
%, with the same perturbation to the homogeneous external field $h^-=-h^+$. 
Again the effect of having a  $-$-constraint in the annulus has a similar effect as imposing $-$-boundary conditions.
%This perturbation is thus small compared to either $h^+$ or $h^-$ so that
Thus for a given $\delta > 0$, e.g. $\delta = \frac{1}{2} M_0(\beta, \alpha)$, 
for  arbitrary $L$ one can find $N(L)$ large enough, such that the expectation of the spin at the origin differs by more than $\delta$. One can therefore  feel the influence from the decimated spins in the far-away annulus,  however large the central interval of decimated alternating spins is chosen. \\
\smallskip
Thus, with our notations, it indeed holds
$M^{+} -M^{-} > \delta$,
uniformly in $L$.

%$\exists 0 < \delta(L) <1$ with
%$$
%M_0^+ > \delta(L) M_0(\beta,\alpha) >0 \; {\rm and} \; M_0^- < - \delta(L) M_0(\beta,\alpha) <0.
%$$
 The essential observation here is that the magnetisations of  Dyson models in an external field are larger in absolute value than those of the $+$ and $-$-measures in zero field, so taking them as boundary conditions everywhere produces the $+$ and $-$-measures. Changing any spins, primed or not, outside $\Delta'$ makes a negligible change when $N(L)$ is chosen large enough, and the Lemma follows, as  choosing  $+$ spins in the annulus produces a magnetisation at the origin of at least $\frac{1}{2} \delta$ and choosing  $-$ spins a magnetisation lower than $- \frac{1}{2} \delta$. 
%Then we are reduced to compare the "complete" magnetisations of the extremal  measures of the Dyson long-range Ising model, and the lemma follows with $\delta = \delta(L) M_0({\beta, \alpha})$ which is strictly positive at low enough temperature for any $1 < \alpha \leq 2$. 
%and long enough order.

$\diamond$ 

Now standard arguments as in \cite{VEFS} provide the non-Gibbsianness.

\section{ Extensions, related issues and comments} 

We have shown that the alternating configuration is a point of essential discontinuity for expectations in the decimation from $\mathbb{Z}$ to $2 \mathbb{Z}$, implying that the associated decimated Gibbs measures are non-Gibbsian. 
In our choice of decimated lattice we 
made use of the fact
%had the advantage
that the constrained system, due to cancellations, again  formed a  
zero-field Dyson-like model. In the case of decimations from $\mathbb{Z}$ to a more diluted lattice $b \mathbb{Z}$ 
%we can do the same, by putting as a constraint alternating intervals of pluses and minuses (or indeed any  configuration of spins which in each even interval alternating with its reflected and spinflipped image. Thus for $b=4$, e.g. the constrained model given as follows $.+--.++-.+--.++-.$ would work).\\
%Other periodic constraints 
the constrained models 
could form ferromagnetic models in a periodically varying external field, with zero mean. 
Although  the original proofs of Dyson  \cite{Dys68} and of Fr\"ohlich and Spencer  \cite{frsP}, or the Reflection Positivity proof of \cite{FILS} do no longer apply to such periodic-field cases, the contour-like arguments of \cite{CFMP} and \cite{Joh} could presumably still be modified to include these cases. Compare also \cite{Ker}.\\

The analysis of \cite{COP} which proves existence of a phase transition for Dyson models in random magnetic fields
for a certain interval of $\alpha$-values should imply that in that case there are many more, random, configurations which all are points of discontinuity. We note that choosing independent spins as a constraint provides a random field which is correlated. However, these correlations decay enough that this need actually not spoil the argument. Similarly, one should be able to  prove that decimation of Dyson models in a weak external field will result in a non-Gibbsian measure.

Estimating the measure of the discontinuity points  leads one to the question of "almost Gibbsian" \cite{MMR}, "intuitively weakly Gibbsian"  \cite{EV} and "weakly Gibbsian" 
properties \cite{MMR}. The analysis of \cite{FP} and  \cite{ALN} extends, due to monotonicity and right-continuity properties, to prove almost Gibbsianness of the transformed measures both with and without a field. This implies as usual (see e.g. \cite{MMR}) weak Gibbsianness with an a.s. convergent  potential as the telescoping one given in \cite{RW}. The latter possesses extra asymptotic properties such as a uniform polynomial decay that should be weaker here. An interesting question would be to perform the analysis of \cite{MRVS}) or \cite{ALN} to get  a.s. configuration-dependent  correlation decays. \\

On the other hand,  the phase transition results of \cite{COP} for the random field Dyson-Ising model, similar to what happens in dimension 3 for the standard n.n. Ising model, strongly indicate that an example of almost surely non-quasilocal transformed measure should be given by the joint measure of this random-field  Dyson-Ising model,  similarly to the 3-dimensional nearest-neighbour random-field Ising model,   following the lines of analysis of \cite{KLNR}. This joint measure then would lack the property of being almost Gibbs and presumably also would violate the variational principle.

%{\bf I think the above sketches answers to the questions below?}\\

%Question 1: are there statements possible about almost and weak Gibbs, along the lines of what Arnaud showed for decimated 2d Ising?\\
%Question 2: extend to decimating bZ instead of 2Z (prove phase transition via Picco's triangle contours?)\\
%Question 3: Is the joint measure of the random-field Dyson model not even almost Gibbs, just as Christof showed for 3d RFIM? Use the results for $\alpha$ close to $\frac{3}{2}$ by Cassandro,Orlandi,Picco?\\

%\section{Comments - Perspectives}
%\subsection{Comments and perspectives}

We have thus extended results which were known before for nearest-neigbour Ising models to a class of long-range models of Dyson type. It turns out that the analogy between varying the dimension and varying the decay parameter of the Dyson models also holds regarding the non-Gibbsianness of various transformed measures, under decimation transformations.
In particular, it turns out that at sufficiently low temperatures the Gibbs measures of the zero-field models, as well as the models in a weak magnetic field under decimation are mapped to non-Gibbsian measures. We expect that, as in the nearest-neighbour case,  the nature of the transformation (decimation, average, majority rule, stochastic evolutions, factor maps...) should not play that much of a role either but we have not pursued our investigations  further in this direction. 
The case of stochastic evolutions (in particular subjecting the Dyson measures to an infinite-temperature evolution)
could  also be investigated,  but may be fairly immediate.  For short times, the results of \cite{LR} imply Gibbsianness for a wide class of evolutions  starting from Gibbs measures with finite-range potentials, and  the effects of the longer ranges of the Dyson-Ising models  should be  be negligible, while  non-Gibbsianness should follow from an analysis more or less along the lines of \cite{EFHR}, and the observations made above, that Dyson models in weak periodic or random fields will have phase transitions at low temperatures, should imply a Gibbs-non-Gibbs transition.

 The fact that long-range models behave analogously to short-range models in higher dimensions as regards their non-Gibbs property is in some sense to be expected. Indeed mean-field models, which have an infinite-range character, show analogous behaviour, as do Kac models which display a long range in a somewhat different fashion \cite{Ku, EnK, KN,ErK, FHM1, FHM2}.  In contrast to the latter, the notion of non-Gibbsianness in the Dyson case is however the same as in the short-range case, no adaptation in its definition is needed. Our proofs also go mostly along the lines of the short-range case, with some modifications due the different proofs of Dyson model phase transitions.

Another class of one-dimensional systems which has attracted a lot of attention over the last years is the class of $g$-measures,  see e.g. \cite{BHR,BK,FM1,Fr}. In the presence of phase transitions, it seems plausible that transforming them also will often map them to non-Gibbsian, cq "non-$g$"-measures. In fact, although it is known that $g$-measures need not be Gibbs measures \cite{FGM, GP}, it appears at this point not known if the Gibbs measures of the Dyson-Ising models can be represented as $g$-measures. \\

 On the other side of the Gibbs-non-Gibbs analysis, when the range of the interaction is lower, i.e. for $\alpha >2$, or the  temperature is too high, uniqueness holds, for all possible constraints and the transformed measures should be Gibbsian.  Some standard high-temperature results apply, which were already discussed in \cite{VEFS}.  About  these shorter-range models,  (i.e. long-range models with faster polynomial decay), Redig and Wang \cite{RW} have proved that Gibbsianness was conserved, providing  in some cases ($\alpha >3$) a decay of correlation for the transformed potential. In our longer-range models,  
for intermediate temperatures 
(below the transition temperature but above the transition temperature of the alternating-configuration-constrained model) decimating both  $+$- and $-$-measures should imply Gibbsianness, essentially due to the arguments as proposed for short-range models in \cite{HK}.\\

%************************************************\\
%I add acknowledgements and try to withdraw a few ref (start with ACCN and BHR)\\
%************************************************\\
 
{\bf Acknowledgments:} We thank S. Friedli and Y. Velenik for valuable remarks and advices, and the referees for their criticisms and comments.

\end{document}